\documentclass[aps,superscriptaddress,floats,showpacs,floatfix,onecolumn]{revtex4}
\usepackage{bm}
\usepackage{graphicx}
\usepackage{subfigure}
\usepackage{amssymb}
\usepackage{amsmath}
\usepackage{color}
\usepackage[a4paper,left=1.5cm, right=1.3cm, top=3.0cm,bottom=2cm]{geometry}

\begin{document}
\title{Transitional boundary layers in low-Prandtl-number convection}
\author{J\"org Schumacher}
\affiliation{Institut f\"ur Thermo- und Fluiddynamik, Technische Universit\"at Ilmenau, Postfach 100565, D-98684 Ilmenau, Germany}
\author{Vinodh Bandaru}
\affiliation{Institut f\"ur Thermo- und Fluiddynamik, Technische Universit\"at Ilmenau, Postfach 100565, D-98684 Ilmenau, Germany}
\author{Ambrish Pandey}
\affiliation{Department of Physics, Indian Institute of Technology, Kanpur, 208016, India}
\author{Janet D. Scheel}
\affiliation{Department of Physics, Occidental College, 1600 Campus Road, M21, Los Angeles, CA 90041, USA}
\date{\today}

\begin{abstract}
The boundary layer structure of the velocity and temperature fields in turbulent Rayleigh-B\'{e}nard flows in closed 
cylindrical cells of unit aspect ratio is revisited from a transitional and turbulent viscous boundary layer perspective. 
When the Rayleigh number is large enough, the dynamics at the bottom and top plates can be separated into an impact region 
of downwelling plumes, an ejection region of upwelling plumes and an interior region away from the side walls. The latter is 
dominated by the shear of the large-scale circulation (LSC) roll which fills the whole cell and continuously varies its orientation. 
The working fluid is liquid mercury or gallium at a Prandtl number $Pr=0.021$ for Rayleigh numbers $3\times 10^5\le Ra \le 4\times 10^8$. 
The generated turbulent momentum transfer corresponds to macroscopic flow Reynolds numbers with
$1.8\times 10^3 \le Re\le 4.6\times 10^4$.
In highly resolved spectral element direct numerical simulations, we present the mean profiles of velocity, Reynolds shear 
stress and temperature in inner viscous units and compare our findings with convection experiments and channel flow data. The complex 
three-dimensional and time-dependent structure of the LSC in the cell is compensated by a plane-by-plane symmetry 
transformation which aligns the horizontal velocity components and all its derivatives with the instantaneous 
orientation of the LSC. As a consequence, the torsion of the LSC is removed  and a streamwise direction in the shear flow 
can be defined. It is shown that the viscous boundary layers for the largest Rayleigh numbers are highly transitional and obey 
properties that are directly comparable to transitional channel flows at friction Reynolds numbers $Re_{\tau}\lesssim 10^2$.
The transitional character of the viscous boundary layer is also underlined by the strong enhancement of the fluctuations of the wall 
stress components with increasing Rayleigh number. An extrapolation of our analysis data suggests that the friction Reynolds number $Re_{\tau}$
in the velocity boundary layer can reach values of 200 for $Ra\gtrsim 10^{11}$. Thus the viscous boundary layer in a liquid metal flow 
would become turbulent at a much lower Rayleigh number than for turbulent convection in gases and gas mixtures.  
\end{abstract}
\maketitle

\section{Introduction}
A better understanding of the local and global mechanisms of turbulent transport of heat and momentum across a fluid layer that is 
heated from below and cooled from above remains a central subject of numerical, theoretical and experimental studies in the field
of turbulent convection \cite{Kadanoff2001,Ahlers2009,Chilla2012}. This setup which is known as the classical Rayleigh-B\'{e}nard 
convection (RBC) case is one ingredient of numerous astro- and geophysical turbulent flows as well as technological applications. A more 
precise quantification of global turbulent transport would immediately improve predictions on structure formation and dynamics. 
The key to deeper insights lies in the boundary layers of the temperature and velocity fields at the top and bottom plates -- understanding 
their transformations with an increase of the temperature difference that is quantified by the Rayleigh number $Ra$.  These boundary layers 
form a bottleneck that limits the transport in fully turbulent convection. The bottleneck is widened when the boundary  layers start to 
fluctuate locally and to become eventually fully turbulent. 

The range of such a transition to boundary layer turbulence would, however, depend strongly 
on the Prandtl number $Pr$ of the convecting fluid which relates viscous to thermal diffusion \cite{Kraichnan1962,Grossmann2000,Grossmann2001}. 
While the thicknesses of the both boundary layers are about the same for $Pr\sim 1$, they differ significantly in the limits of very small and very large 
Prandtl number. Their dynamics are then more loosely coupled since one of the layers is well embedded in the dissipation-dominated 
and spatially smooth sublayer of the other field. For liquid metal convection at $Pr\ll 1$ the thermal boundary layer is much thicker than the viscous 
boundary layer, such that the latter is  well embedded in the diffusive sublayer where temperature decreases to a good approximation linearly with 
respect to wall distance \cite{Siggia1994,Chung1992}.  This opens the possibility to disentangle their dynamics and to compare the statistics of the viscous
boundary layer to standard turbulent wall-bounded flows without temperature differences. 

The limit of very low Prandtl number convection is interesting for a further reason.  In ref. \cite{Schumacher2015}
it was shown recently that the highly diffusive temperature field and the resulting coarse plumes drive the fluid turbulence more vigorously
than the more filamented plumes at larger Prandtl and comparable Rayleigh number. Additional studies in \cite{Scheel2016} found that the same
holds for the boundary layer of the velocity field. The level of the local fluctuations and the turbulent drag are enhanced in line with a significantly increased
global momentum transfer which is measured by the Reynolds number $Re$. In this way, a low-Prandtl-number convection flow at a given 
Rayleigh number will obey a much more vigorous fluid turbulence than a convection flow 
in air or water and thus provide an appropriate setup to investigate the transitional character of the (velocity) boundary layer in detail.  

In the present work, we study the boundary layer dynamics by means of high-resolution direct numerical simulations (DNS) which can
access all details of the fluctuating turbulent fields in the RBC flow. The setup that is chosen agrees with one of the most 
common laboratory experiments: a closed cylindrical cell with an aspect ratio of one. Compared to a cubical or rectangular cell, this setup
sustains one statistically homogeneous coordinate in the system, the azimuthal one, and has thus the highest symmetry. 

The perspective that is taken here is to analyse 
our simulation data as for a viscous boundary layer in a pressure-driven channel flow with a uni-directional mean flow \cite{Smits2011}. The 
high-Rayleigh-number convection in the closed cylindrical cell builds up a large-scale circulation (LSC) which changes its orientation in the course of the 
dynamical evolution \cite{Ahlers2009,Shi2012}.  For a better comparison to a channel flow, streamwise and spanwise directions will be obtained 
in the convection case by a plane-by-plane rotation of the velocity field into the horizontal direction of the LSC. This symmetry transformation 
removes the torsion in the large-scale 
circulation. In low-Prandtl-number convection this circulation roll turns out to perform a very coherent motion since it is driven by coarse thermal 
plumes. Mean profiles of the streamwise velocity, the temperature and the Reynolds stresses are analysed in inner
wall units. Therefore, we have to adapt definitions of the friction velocity and the friction temperature to the present setup.  Furthermore, the 
statistics of the wall-normal derivatives of the horizontal velocity components is compared to those of the channel flow. The main motivation of the present
work is to better quantify the transitional character of the boundary layers in the RBC flow.   

The outline of the manuscript is as follows. In the next section, the equations of motion and some details on the numerics are 
given. Section \ref{symmetry} analyses the large-scale flow and presents the symmetry transformations. Section \ref{means} lists
our results for the mean profiles of temperature, Reynolds shear stress  and streamwise velocity. Additionally, we derive skin friction Reynolds numbers for the 
individual runs of our data record. Section \ref{derivatives} summarizes our findings for the derivatives 
at the wall. Finally all results are summarized.

\begin{figure*}
\begin{center}
\includegraphics[scale=0.13]{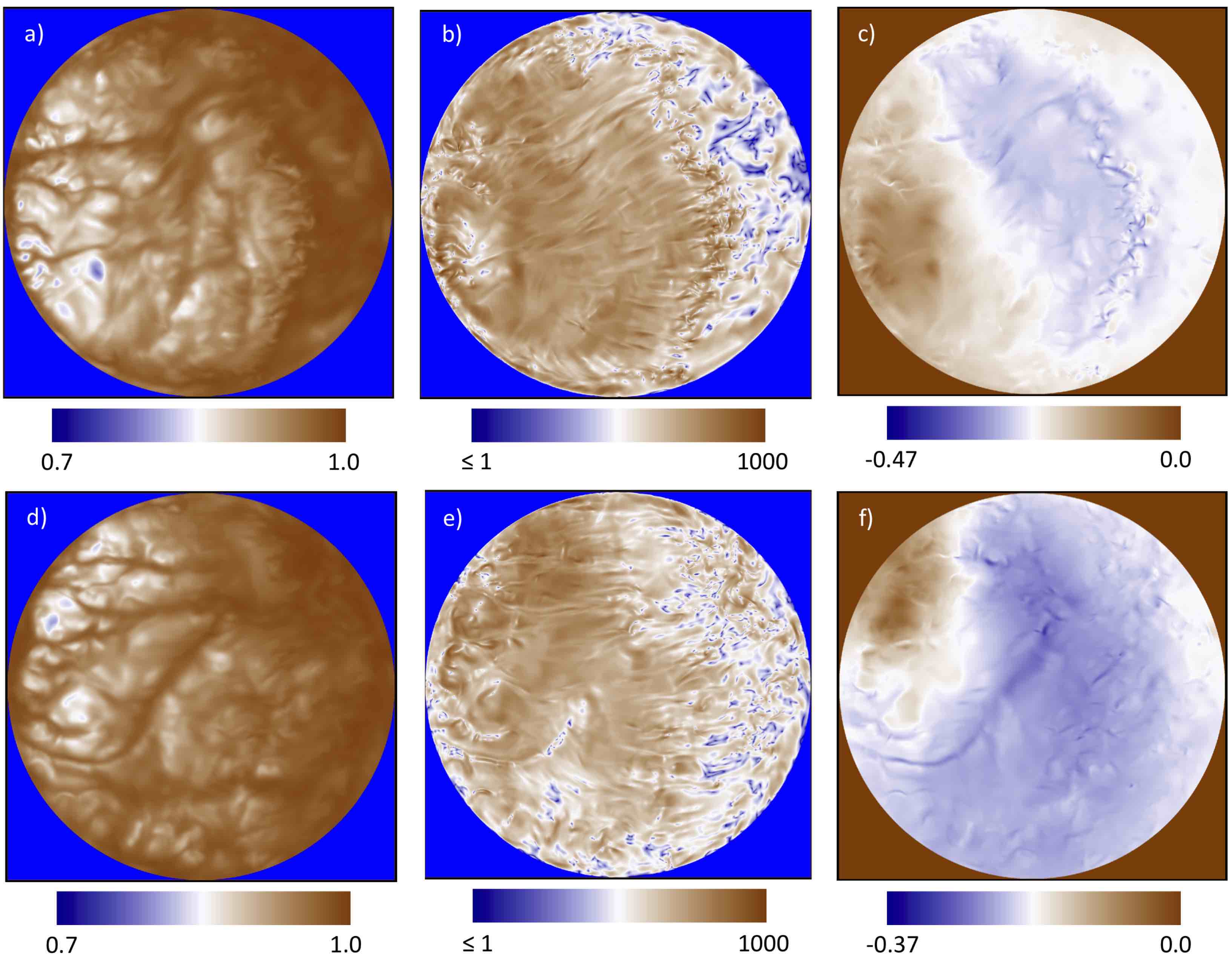}
\caption{Boundary layer structure in a turbulent convection flow at $Pr=0.021$ and $Ra=10^8$. Two snapshots are shown 
at $t=35.1$ (upper row) and $37.1$ (lower row). (a,d) Temperature $T$ at $z=0.0024$ which corresponds with $0.09 \delta_T$.
(b,e) Magnitude of skin friction $|{\bm s}|$ in logarithmic units (see equation (\ref{wall})) at $z=0$. (c,f) Pressure $p$ at  $z=0.0024$. 
The view of the bottom plate is from below.}
\label{Fig1}
\end{center}
\end{figure*}

\section{Simulation model and parameters}
\label{methods}
\begin{figure}
\begin{center}
\includegraphics[scale=0.12]{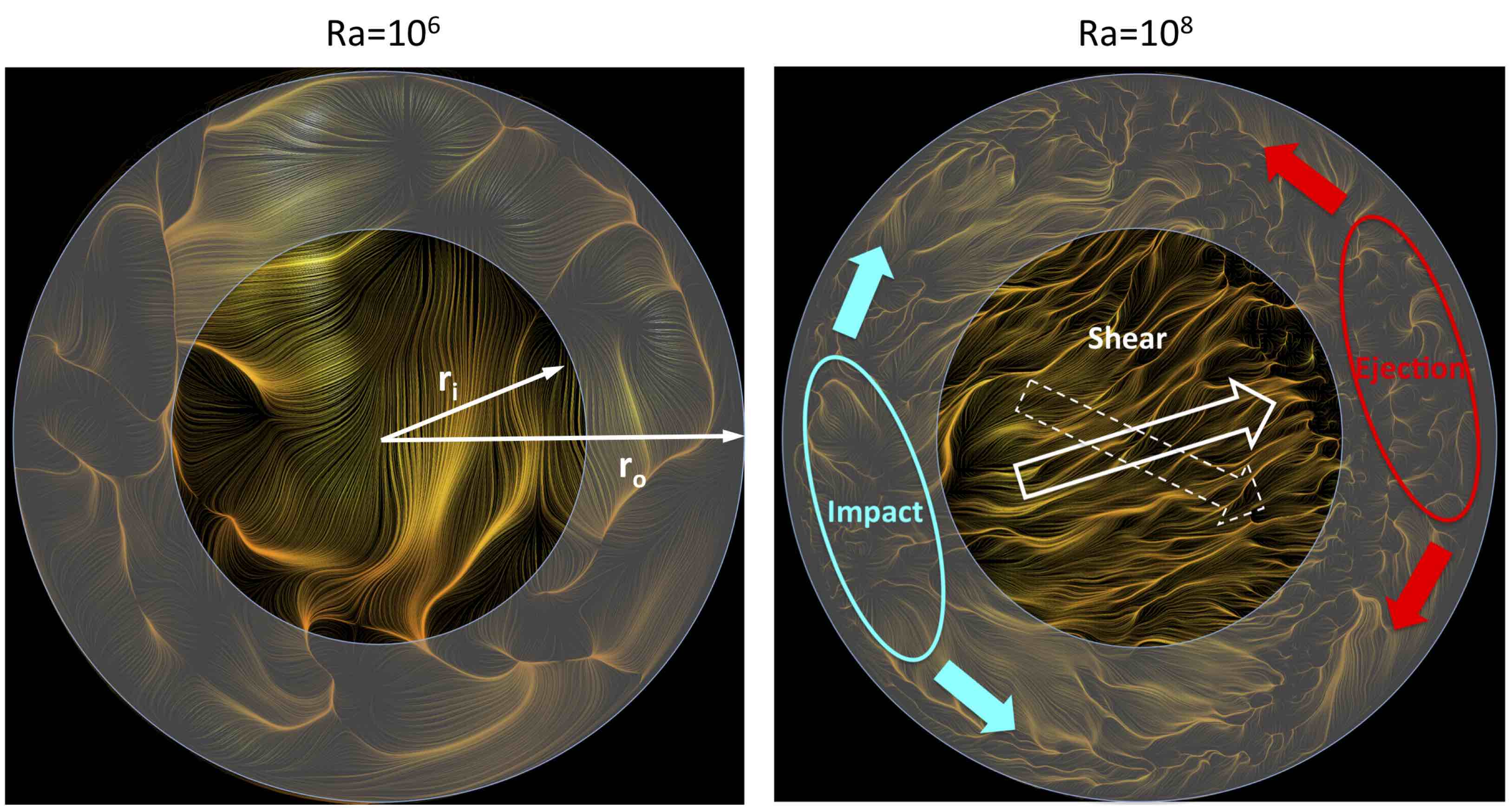}
\caption{Boundary layer structure in a turbulent convection flow for RBC 2 (left) and RBC4 (right). Streamlines of the skin friction 
field at the bottom plate (see equation \ref{wall}). Impact, shear and ejection sections are indicated. The block arrows in all three sections indicate the 
temporal variations. The interior plate section for data points with $r\le r_{\text i}$ is highlighted and will be used for most of the analysis. 
The view on the bottom plate is from below as in figure \ref{Fig1}.}
\label{Fig2}
\end{center}
\end{figure}

We solve the three-dimensional equations of motion in the Boussinesq approximation. The equations are made dimensionless by using the
height of the cell $\tilde{H}$, the free-fall velocity $\tilde{U}_f=(\tilde{g} \tilde{\alpha} \Delta \tilde{T} \tilde{H})^{1/2}$ and the imposed temperature difference $\Delta \tilde{T}$. Times are measured in free-fall time units $\tilde{T}_f=\tilde{H}/\tilde{U}_f$. (Quantities with a physical dimension
are given with a tilde.) The equations 
contain the three control parameters: the Rayleigh number $Ra$, the Prandtl number $Pr$ and the aspect ratio $\Gamma=2\tilde{r}_{\text{o}}/\tilde{H}$ 
with the cell radius $\tilde{r}_{\text{o}}$ (see figure \ref{Fig2}). The set of dimensionless equations is given by
\begin{eqnarray}
\label{ceq}
 {\bm \nabla}\cdot {\bm u}&=&0\,,\\
\label{nseq}
\frac{\partial{\bm u}}{\partial  t}+({\bm u}\cdot{\bm\nabla}){\bm u}
&=&-{\bm \nabla}  p+\sqrt{\frac{Pr}{Ra}} {\bm \nabla}^2{\bm u}+  T {\bf e}_z\,,\\
\frac{\partial  T}{\partial  t}+( {\bm u}\cdot {\bm \nabla})  T
&=&\frac{1}{\sqrt{Ra Pr}} {\bm \nabla}^2  T\,,
\label{pseq}
\end{eqnarray}
where
\begin{equation}
Ra=\frac{\tilde{g}\tilde{\alpha}\Delta \tilde{T} \tilde{H}^3}{\tilde{\nu}\tilde{\kappa}}\,,\;\;\;\;\;\;\;\;Pr=\frac{\tilde{\nu}}{\tilde{\kappa}}\,.
\end{equation}
The variable $\tilde{g}$ stands for the  acceleration due to gravity, $\tilde{\alpha}$ is the thermal expansion coefficient, $\tilde{\nu}$ is the kinematic viscosity, and 
$\tilde{\kappa}$ is thermal diffusivity. We use an aspect ratio of  $\Gamma=1$ here. No-slip boundary conditions for the fluid (${\bf u}=0$)  are applied at 
the walls. The side walls are thermally insulated ($\partial T/\partial {\bm n}=0$) and the top and bottom plates are held at constant dimensionless
 temperatures $T=0$ and 1, respectively. In response to the input parameters $Ra$, $Pr$ and $\Gamma$,  turbulent heat and momentum fluxes 
 are established. The turbulent heat transport is determined by the Nusselt number which is defined as
\begin{equation}
Nu=\frac{\tilde{Q} \tilde{H}}{\tilde{\kappa} \Delta\tilde T}\;\;\;\text{with}\;\;\;
\tilde{Q}=\langle \tilde{u}_z\tilde{T}\rangle_{A,t}-\tilde{\kappa}\Bigg\langle\frac{\partial \tilde T}{\partial \tilde z}\Bigg\rangle_{A,t}\,,
\label{Nusselt1}
\end{equation}
 with an area-time average $\langle\cdot\rangle_{A,t}$. Note that $Q$ is a constant in each horizontal cross section $A$.
 Equation (\ref{Nusselt1}) can be rewritten as
\begin{equation}
Nu=1+\sqrt{Ra Pr}\langle  u_z  T\rangle_{V,t}\,,
\label{Nusselt2}
\end{equation}
with a volume-time average $\langle\cdot\rangle_{V,t}$.
The turbulent momentum transport is expressed by the (large-scale) Reynolds number which is defined as
\begin{equation}
Re=u_{rms,V}\sqrt{\frac{Ra}{Pr}}\;\;\;\;\text{with}\;\;\;\;u_{rms,V} =\sqrt{\langle u_i^2\rangle_{V,t}}\,.
\label{Reynolds}
\end{equation}

The equations are numerically solved by the Nek5000 spectral element method package which has been 
adapted to our problem. The code employs second-order time-stepping, using a backward difference formula. The whole set of
equations is transformed into a weak formulation and discretized with a particular choice of spectral basis functions 
\cite{Fischer1997,Deville2002}. For further numerical details and comprehensive tests of the sufficient spectral resolution, 
we refer to detailed investigations in \cite{Scheel2013}. 

The cylindrical cell is resolved by up to 6.27 million spectral elements and the spectral expansion of all turbulent fields is done 
with Lagrangian interpolation polynomials up to order 13 in each spatial direction which results in a $14^3$ collocation grid on 
each spectral element. The simulation run at the largest Rayleigh number was conducted on 524,288 MPI tasks of the Blue Gene/Q 
system Mira at Argonne National Laboratory. The time advancement of 6 free-fall times took about 50 million core hours. 

We focus on five data sets (see Table \ref{Tabpran}) for Rayleigh-B\'{e}nard convection in liquid mercury at $Pr=0.021$ which are 
denoted by RBC1 to RBC5 and cover more than three orders of magnitude in terms of the Rayleigh number, $3\times 10^5\le Ra\le 4\times 10^8$.  
For a large fraction of the paper we study in detail a sequence of snapshots for RBC4 over a time span of $6.7 T_f$ which are separated 
by approximately $0.12 T_f$ from each other.  

An additional DNS for a pressure gradient driven channel flow (CF) is used for comparison. It is based on a finite difference method
with uniform grid spacing in the horizontal directions with periodic boundaries and with a non-uniform grid in $z$-direction that 
corresponds with the Chebyshev collocation points \cite{Krasnov2011}. The channel has the extensions $L_x:L_y:L_z=4\pi:2\pi:2$.

Results for $Ra = 1\times 10^8$ and Pr=$0.021$ (RBC4) are shown in Figure \ref{Fig1}. Temperature,  magnitude of the skin friction field 
(see Section \ref{symmetry} for definition), and 
pressure are shown for horizontal slices through the bottom boundary layer, for two different instants in time. One sees that the temperature 
is very diffuse. One also sees the overall large-scale direction of the flow, which changes with time from approximately 0.7 radians in the top 
plot to 6 radians in the bottom plot. Finally the pressure shown in Figure \ref{Fig1}(c,f) has a fairly steep favorable gradient near the impact 
region, but then becomes fairly flat and then rises slightly in the ejection region.
  
\section{Symmetry-breaking large-scale flow}
\label{symmetry}
It is known that the large-scale circulation (LSC) in a closed cylindrical convection cell has a complex three-dimensional 
structure \cite{Brown2006,Xi2008,Wagner2012,Shi2012}. For aspect ratio $\Gamma=1$, the wind, which is averaged over  
6-30 free-fall times $T_f$, takes the form of a single flow roll with a preferred orientation -- a configuration that clearly breaks azimuthal symmetry. 
This roll is additionally twisted and changes orientation slowly in time.  It is thus expected that statistical homogeneity in the azimuthal direction 
can be re-established for a very long time interval only. First estimates in \cite{Emran2015} suggest times $t\gtrsim 10^4 T_f$ or 
even larger. Statistical sampling can typically be done in DNS over shorter time intervals only, particularly for simulation runs at the highest Rayleigh 
numbers.

Figure \ref{Fig2} displays instantaneous snapshots of the streamlines of the two-dimensional skin friction vector field at the
bottom plate for convection in mercury at $Ra=10^6$ (left) and $10^8$ (right). The skin friction field can be considered as
a blueprint of the near-wall viscous boundary layer  dynamics and has been studied in wall-bounded shear flow \cite{Chong2012,Lenaers2014}
as well as in Rayleigh-B\'{e}nard convection \cite{Bandaru2015}. At the bottom plate ($z=0$), the velocity gradient tensor $\tilde{A}_{ij}$ takes 
the following form
\begin{equation}
\label{tensor}
  \hat{A}\Big|_{z=0} = \left(\begin{array}{ccc}
                                0 & 0 & \partial \tilde{u}_x/\partial \tilde{z} \\
                                0 & 0 & \partial \tilde{u}_y/\partial \tilde{z} \\
                                0 & 0 & 0
                                \end{array}\right)\:.
\end{equation}
Both components form a two-dimensional wall shear stress vector field and a related skin friction field. They are defined as
\begin{equation}
  \label{wall}
\tilde{\bm \tau}_w=\tilde{\rho}_0 \tilde{\nu} \frac{\partial \tilde{{\bm u}}^{(2)}}{\partial \tilde z}\Bigg|_{\tilde z=0}\,,
\quad\quad\text{and}\quad\quad\tilde{\bm s}=\frac{\tilde{\bm \tau}_w}{\tilde{\rho}_0 \tilde{\nu}}\,.
\end{equation}
The superscript denotes the two horizontal (or tangential) $x$-- and $y$--components. Particularly for the higher Rayleigh number, 
one can clearly divide the near--plate boundary layer into three main regions, the 
{\em impact} region where the cold LSC flow masses hit the bottom plate, the {\em shear} region where the LSC sweeps across 
the interior section of the plate, and the {\em ejection} region where the heated fluid rises up towards the cold top plate again. This 
separation into three distinct regions requires a sufficiently large Rayleigh number. We will return to this point in Section V when 
discussing the derivatives at the plates at $z=0$ and 1. On the basis of the critical points of the skin friction field the inner
region can be clearly distinguished from the impact and ejection regions. Also visible is the broken azimuthal symmetry 
of the flow. 

We define two different area--time averages, one across the whole plate $A$ with $r\le r_{\text o}=0.5$ which will be denoted by 
$\langle f\rangle_{A,t}$ (as already mentioned in section \ref{methods}) and one across an interior section of the plate which is indicated in 
figure \ref{Fig2} for points with $r\le r_{\text i}$. If not stated otherwise, $r_{\text{i}}=0.3$ is taken.
The latter will be denoted as $\langle f\rangle_{b,t}$.  As seen in figure \ref{Fig2}, the average with respect to the interior plate
section excludes impact and ejection region and brings us closest to the conditions in a canonical boundary layer with a unidirectional
mean flow, at least for the higher Rayleigh number. Therefore, most of the statistical analysis is restricted to this inner region in the following.

\begin{figure}
\begin{center}
\includegraphics[scale=0.3]{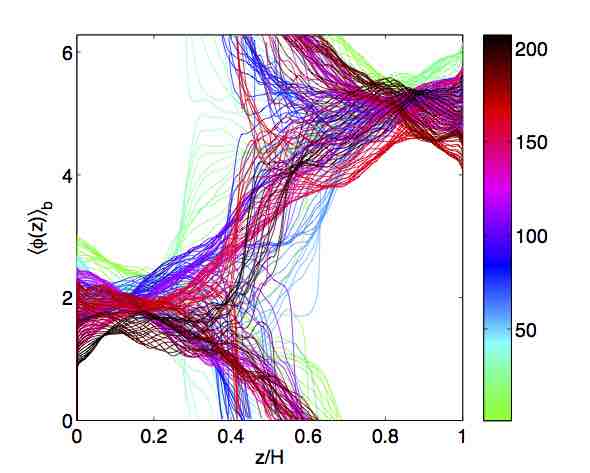}
\includegraphics[scale=0.3]{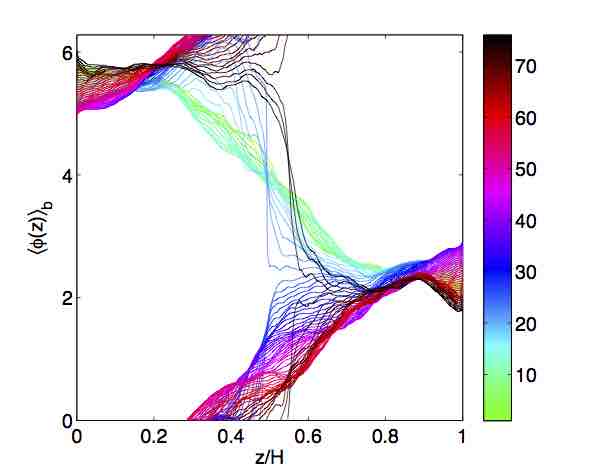}
\caption{Local orientation angle $\langle \phi(z)\rangle_b$ (see equation (\ref{angle})) as a function of $z/H$ for consecutive snapshots for RBC3 (left) and RBC5 (right). The color bar coding is for snapshot number and each snapshot is output at regular spaced intervals of time, $0.14 T_f$ for RBC3  and $0.08 T_f$ for RBC5. Note
that the angle is not defined at the plates due to no-slip boundary conditions.}
\label{phivsz}
\end{center}
\end{figure}

The local orientation angle is also calculated in each plane at fixed height $z>0$ by 
\begin{equation}
\label{angle}
\langle {\phi}(z,t)\rangle_b = \arctan \left [\frac{\langle u_y(z,t)\rangle_{b}}{\langle u_x(z,t)\rangle_b} \right]\,.
\end{equation}
As indicated by the filled arrows in the right panel of figure \ref{Fig2}, impact and ejection regions will slowly move azimuthally. We investigate this further in figure \ref{phivsz}, where we show instantaneous profiles of $\langle\phi(z)\rangle_b$  for two different Rayleigh numbers. One sees that the orientation angle twists as $z$ increases from the angle at $z\simeq 0$ , to eventually match the orientation angle at $z\simeq 1$ (which is different from the angle at $z\simeq 0$ 
by about $\pi$ radians). If we focus on the right panel of figure \ref{phivsz}, one sees that sometimes this twist is clockwise (as for the first 15 snapshots), other times it is counterclockwise (as for snapshots 30-60) and sometimes the twist changes direction (as seen in  snapshots 20-30 and the last three as well). Of course, near the center of the container the horizontal velocity is significantly reduced compared to near the top and bottom plates, but we still see a steady twist in $\langle \phi(z)\rangle_b$ for most of the time, even in the center of the container. A similar behavior is seen for the left panel of figure \ref{phivsz}, although the behavior occurs more rapidly in these $T_f$ time units.

\begin{figure}
\begin{center}
\includegraphics[scale=0.3]{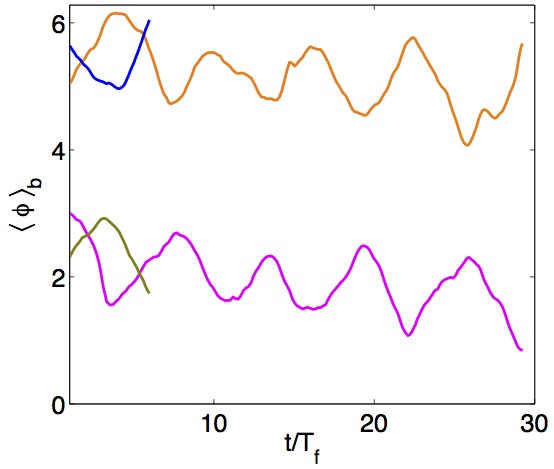}
\includegraphics[scale=0.3]{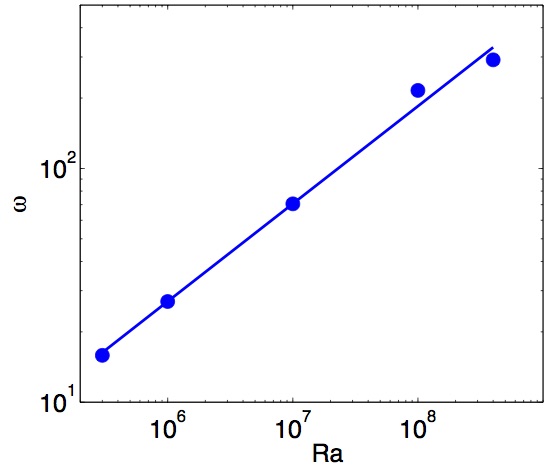}
\caption{Left panel: Local orientation angle $\langle \phi\rangle_b$ (see equation (\ref{angle})) as a function of time as measured in free-fall time units $T_f$ for $z\simeq 0$ (bottom plate) and $z \simeq 1$ (top plate) for RBC3  (magenta = bottom and orange = top ) and  RBC5 (blue =bottom and grey = top). Right panel: Oscillation frequency $\omega$  of the local orientation angle $\langle \phi\rangle _b$ as a function of $Ra$.  The line is a fit to the data and gives $(0.08 \pm 0.05) Ra^{0.42\pm 0.02}$. The frequency $\omega$ is in radians per diffusive time units $\tilde{t}_d = \tilde{H}^2/\tilde{\kappa}$.}
\label{phivst}
\end{center}
\end{figure}

One also sees that the local orientation angle for fixed $z$ oscillates with time. The angle $\langle \phi \rangle_b$ is plotted near the bottom ($z\simeq0$)  
and top ($z \simeq 1$) of the container in the left panel of figure \ref{phivst} for the two representative cases, RBC3  and RBC5 as a function of the time $t/T_f$. 
For both Rayleigh numbers we see the angle switches or oscillates, with the angle at the bottom plate  out of 
phase with the angle at the top plate. We measure the frequency of these oscillations $\omega$ and plot this versus $Ra$ in the  right panel of figure \ref{phivst}. 
This oscillation frequency is measured in units of radians per (dimensionless) diffusive time units $t_d$. One can convert from free-fall time units 
$\tilde{T}_f$ to diffusive time units $\tilde{t}_d$ by $\tilde{t}_d=\sqrt{Ra\, Pr}\, \tilde{T}_f$.
The oscillation frequency increases with $Ra$ which is in agreement with previous results for $Pr \approx 0.021$, $5\times 10^5 < Ra < 5\times 10^9$ \cite{Cioni1997} as well as for $Pr = 6,  7\times 10^7 < Ra < 3\times 10^9$  \cite{Funfschilling2004} and $Pr=19.4, 8 \times 10^8 < Ra < 3\times 10^{11}$ \cite{Xie2013}, all at $\Gamma = 1$. The exponent of the fit of $\omega$ versus $Ra$ is $0.42 \pm 0.02$ which agrees remarkably well with the exponent of $0.424$ of \cite{Cioni1997}. The experiments of \cite{Funfschilling2004} and  \cite{Xie2013} measured an exponent of $0.36$ which is a bit lower. Also the magnitude of the oscillation frequencies that we measured for $Pr=0.021$ are lower than those measured for $Pr=6$ by a factor of 30, indicating that lower Prandtl number stabilizes the  oscillations of the LSC for a given $Ra$.

How can the mean velocity profile be determined under such circumstances? The definitions which are applied in the theory of 
classical turbulent boundary layers after a Reynolds decomposition use  streamwise and spanwise directions.
In contrast to a canonical boundary layer or a wall-bounded flow, a proper mean flow determination in RBC has to be adjusted to 
these permanently changing conditions. One has to determine a mean horizontal wind orientation for each plane at a given height 
$z$ and for each time instant.

\begin{figure}
\begin{center}
\includegraphics[scale=0.4]{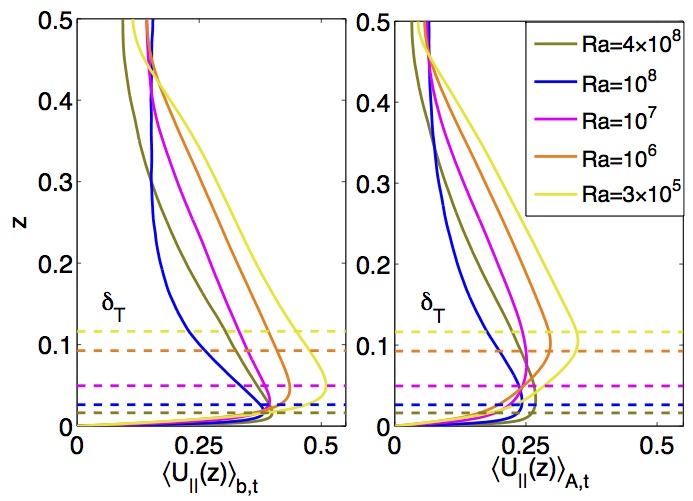}
\caption{Vertical profiles of $\langle U_{\parallel}(z)\rangle$. Left: Profiles averaged over the interior region. 
Right: Profiles averaged over the whole plate. The corresponding boundary layer thicknesses $\delta_T=1/(2 Nu)$ are indicated by 
horizontal lines with the same color. In all cases the profiles taken from the top and bottom plate are included in the time average.}
\label{prof1}
\end{center}
\end{figure}

A planar rotation $\hat{R}_3(z,t)$ by $\langle {\phi}(z,t)\rangle_b$ defines a new coordinate frame that is aligned in each plane $z$ and at each time $t$ 
with the mean wind direction above the interior section $b$ with $r\le r_i$ or the full plate. New coordinates and velocity components are then 
given by 
\begin{equation}
  \label{rot}
  \left(\begin{array}{c}
                                x_{\parallel} \\
                                y_{\parallel} \\
                                z_{\parallel}                                
                                \end{array}\right) = 
   \left(\begin{array}{ccc}
                                \cos \langle \phi\rangle_b  & \sin \langle \phi\rangle_b  & 0 \\
                                -\sin \langle \phi\rangle_b  & \cos \langle \phi\rangle_b  & 0 \\
                                0 & 0 & 1
                                \end{array}\right) 
  \left(\begin{array}{c}
                                x \\
                                y \\
                                z                                
                                \end{array}\right) \;\;\;\;\mbox{and}\;\;\;\;
  \left(\begin{array}{c}
                                U_{\parallel} \\
                                V_{\parallel} \\
                                W_{\parallel}                                
                                \end{array}\right) = 
   \left(\begin{array}{ccc}
                                \cos \langle \phi\rangle_b  & \sin \langle \phi\rangle_b  & 0 \\
                                -\sin \langle \phi\rangle_b  & \cos \langle \phi\rangle_b  & 0 \\
                                0 & 0 & 1
                                \end{array}\right) 
  \left(\begin{array}{c}
                                u_x \\
                                u_y \\
                                u_z                                
                                \end{array}\right) \;.               
\end{equation}
The rotated velocity components define the new streamwise ($U_{\parallel}$), 
spanwise ($V_{\parallel}$) and wall-normal components, respectively. 
The area--time averages of the streamwise component $\langle U_{\parallel}(z)\rangle_{A,t}$ and $\langle U_{\parallel}(z)\rangle_{b,t}$ are 
shown in figure \ref{prof1}. We verified that the spanwise mean, $\langle V_{\parallel}(z)\rangle_{A,t}$ and $\langle V_{\parallel}(z)\rangle_{b,t}$ 
are now indeed zero across the whole height. As expected, the restriction to the plate interior leads to an increase of the amplitude of the mean 
streamwise velocity which is visible by a comparison of the left and right panels of figure \ref{prof1}.  Furthermore, the maxima of the mean profile for the plate interior (left panel of figure \ref{prof1}) are always closer to the wall which indicates a smaller local boundary thickness in the interior. This is in 
agreement with \cite{Scheel2014,Scheel2016}. Note that the profiles for $Ra=4\times 10^8$  in both panels of figure \ref{prof1} 
do not quite follow the trends as for the rest of the Rayleigh numbers. This is because this simulation could not be run for as long, and hence fewer 
statistics were gathered.

To summarize this section, this planar rotation has brought the complex large scale flow
closest to a standard boundary layer case. We have removed the torsional degrees of freedom from the flow. A similar (not the same) 
idea was investigated for a plane Poiseuille flow with periodic boundary conditions in the streamwise and spanwise directions by Kreilos 
et al. \cite{Kreilos2014}.    

\section{Mean profiles in the boundary layer}
\label{means}
\subsection{Mean streamwise velocity}
\begin{figure}
  \begin{center}
\includegraphics[scale=0.3]{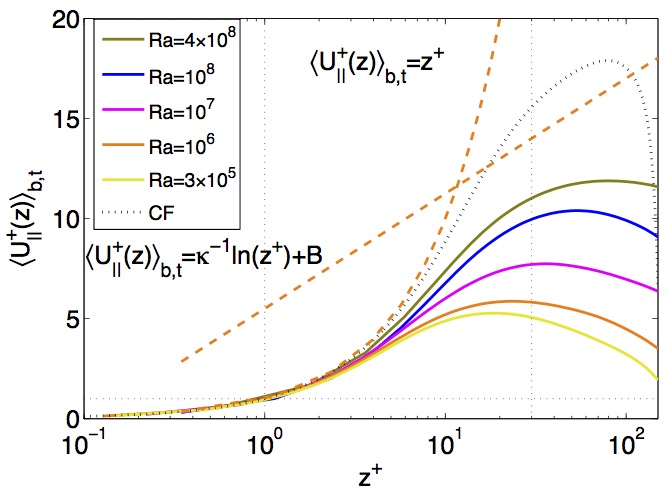}
\includegraphics[scale=0.3]{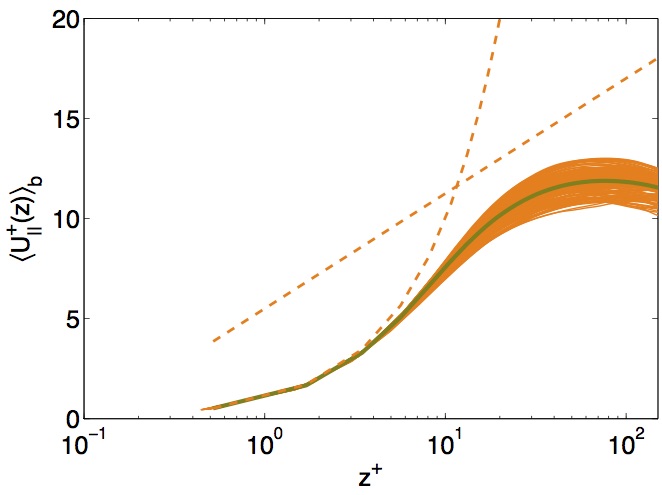}
\caption{Left: Semi-logarithmic plot 
of the vertical profiles of $\langle U^+_{\parallel}(z)\rangle_{b,t}$ versus $z^{+}$. The linear law in the viscous buffer layer and the 
logarithmic law of the wall are also indicated. The von K\'{a}rm\'{a}n constant is $\kappa=0.4$ and the offset is $B=5.5$. For comparison,
we also plot a profile which is obtained in a channel flow simulation at the same friction Reynolds number as the run with the highest 
Rayleigh number. Right: Velocity profiles for RBC5. The grey line is the same as in the left panel.  The instantanous profiles for all 75 snapshots (in orange) 
are also plotted here, to give a sense of the range of variation of such profiles with time. In all RBC cases the profiles taken from the top and bottom plate are included.}
\label{mean1}
\end{center}
\end{figure}

As a next step, we now study how the mean streamwise velocity compares to a turbulent boundary layer. 
The dimensionless friction velocity is given by 
\begin{equation}
\label{flux4}
u_{\tau}= \left(\frac{Pr}{Ra}\right)^{1/4} \Biggl\langle\left(\Bigg\langle\frac{\partial u_x}{\partial z} 
\Bigg\rangle^2_{b}+\Bigg\langle\frac{\partial u_y}{\partial z}
\Bigg\rangle^2_{b}\right)^{1/4}\Bigg|_{z=0}  \Biggl\rangle_t\,.
\end{equation}
The rotation (\ref{rot}) is not defined at $z=0$ since both velocity components are exactly zero in (\ref{angle}). 
Thus one is left with the original wall-normal derivatives at the plate.
And a similar equation is used for the top plate except the gradients are evaluated at $z=1$. Note that (\ref {flux4})  is similar to 
the equation (3.8) in \cite{Scheel2014}. The viscous length scale of a turbulent boundary layer is given by $\tilde{z}_{\tau}= 
\tilde\nu/\tilde u_{\tau}$. The dimensionless length is then given by 
\begin{equation}
\label{flux5}
z_{\tau}=\sqrt{\frac{Pr}{Ra}} \, u_{\tau}^{-1}\,.
\end{equation}
Figure \ref{mean1} (left) shows the logarithmic velocity profile, $\langle U^+_{\parallel}(z)\rangle_{b,t}=
\langle\langle U_{\parallel}(z,t)\rangle_{b}/u_{\tau}(t)\rangle_t$ versus $z^+=\langle z/z_{\tau}(t) \rangle_t$. Specifically, 
the bulk-averaged instantaneous  logarithmic velocity profiles are scaled with each individual $u_{\tau}(t), z_{\tau}(t)$ 
and then time-averaged. This provides a more dynamic estimate of the profiles, similar to what was done in \cite{Zhou2010}.
We also indicate the linear scaling in the viscous sublayer which is well resolved in our DNS and a logarithmic law of the wall for a canonical 
turbulent velocity boundary layer with the standard von K\'{a}rm\'{a}n constant $\kappa = 0.4$ and offset coefficient $B = 5.5$. 

It is seen that the profiles do approach the logarithmic law as $Ra$ increases, but they are not yet turbulent enough to reach the 
canonical log law. Finally for comparison, a profile is plotted which is obtained in a channel flow simulation at the 
same friction Reynolds number as the run with the highest Rayleigh number (see table \ref{Tabpran}). Interestingly the channel flow comparison plot shows 
an overshoot which is typical in channel flow for  Reynolds numbers that are too low to be turbulent in the sense that they follow the 
logarithmic law \cite{Iwamoto2002,Elsnab2011}. However, this is not true for the RBC case, where the profiles are consistently below the log law.

To obtain a sense of the uncertainly in calculating these profiles, the same time-averaged profile is plotted for RBC5 as the green curve 
in the right panel of figure \ref{mean1} along with all 75 instantaneous profiles in orange. One does see these curves 
instantaneously approaching even closer to the logarithmic law, revealing the transitional nature of these boundary layer profiles.

The friction Reynolds number is defined here as $Re_{\tau}=\tilde{u}_{\tau} \tilde{\delta}_{\ast}/\tilde\nu$. In our scaled units this translates to
\begin{equation}
\label{Re1}
Re_{\tau}=u_{\tau}\delta_{\ast}\,\sqrt{\frac{Ra}{Pr}}\,.
\end{equation}
The relevant length scale used here is the $z$ position of the maximum of the time-averaged profile and is denoted as $\delta_{\ast}$ and 
scaled in units of $H$. Note that we use the time-averaged profile instead of the maximum of each instantaneous profile, since there is too 
much variability in local profiles for the instantaneous method to always provide a well-defined $\delta_{\ast}(t)$.  We do still use our local 
$u_{\tau}(t)$ which enables us to estimate the error bars associated with $\langle Re_{\tau}(t) \rangle_t$.

\begin{figure}
\begin{center}
\includegraphics[scale=0.4]{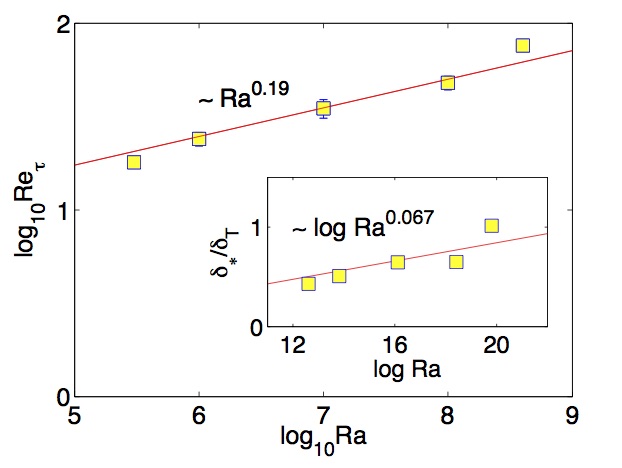}
\caption{Friction Reynolds number $Re_{\tau}$ versus Rayleigh number $Ra$. The red line is a power law fit 
to the data and gives $Re_{\tau}=(1.75 \pm 0.3)\times Ra^{0.19 \pm 0.01}$. Inset: Ratio of maximum of the streamwise velocity profile 
$\delta_{\ast}$ to the thermal boundary layer thickness $\delta_T$. The red line is a fit to the data, $\delta_{\ast}/
\delta_T=(0.067 \pm 0.02) \log(Ra)-(0.4 \pm 0.3)$. Note that most of the error bars are too small to be seen.}
\label{ratio}
\end{center}
\end{figure}

In table \ref{Tabpran} both Reynolds numbers are listed for all simulation runs. The magnitudes of $Re_{\tau}$ consistently take values for 
which a turbulent boundary layer is not yet established in a canonical channel flow which are $Re_{\tau}\lesssim 200$
\cite{Iwamoto2002,Kim1987}. In figure \ref{ratio} we plot the friction Reynolds number versus Rayleigh number
and detect for the range of Rayleigh number an approximate growth as a power law. The inset of the figure displays
the ratio of $\delta_{\ast}$, the distance from the wall at which the maximum streamwise velocity in the interior section is found, to $\delta_T$, 
the thermal boundary layer thickness. This distance is steadily increasing towards one which can be interpreted
as a growth of the velocity bursts. Finally using the fit in figure \ref{ratio} we estimate that the Rayleigh number at which $Re_{\tau} =  200$ 
is $Ra = (1 \pm 5)\times 10^{11}$ for $Pr=0.021$.
\begin{table}[t]
\begin{center}
\begin{tabular}{ccccccc}
\hline\hline
Run & $Ra$ &  $Pr$ & $u_{rms}$ & $Re$ & $Re_{\tau}$ & $2r_{\rm i}/z_{\tau}$ \\ 
\hline
RBC1 & $3\times 10^5$  & 0.021 & 0.483 $\pm$ 0.009 & 1830 $\pm$ 30  & 18$\pm$1 & 220 $\pm$ 12 \\
RBC2 & $10^6$  & 0.021 & 0.439 $\pm$ 0.006 & 3030 $\pm$ 40  & 24$\pm$ 2 & 300 $\pm$ 30 \\
RBC3 & $10^7$  & 0.021 & 0.387 $\pm$ 0.005 & 8450 $\pm$ 100  & 35$\pm$4 & 650 $\pm$ 80\\
RBC4 & $10^8$  & 0.021 & 0.332 $\pm$ 0.004 & 22900 $\pm$ 300 & 48 $\pm$ 4 & 1700 $\pm$ 130 \\
RBC5 & $4\times 10^8$ & 0.021 & 0.334 $\pm$ 0.004  & 46000 $\pm$ 600  & 76 $\pm$ 5 & 2800 $\pm$ 190 \\
CF & -- & -- &  1.054  &  1145  &  78  &  989\\
\hline\hline
\end{tabular}  
\caption{Parameters of the five different spectral element simulations RBC1 to RBC5 and the channel flow simulation CF. 
The root mean square velocity is obtained as a space-time average over the whole cell volume. The large-scale Reynolds number 
is defined by (\ref{Reynolds}) and friction Reynolds number by (\ref{Re1}). Finally, the ratio $2 r_{\rm i}/z_{\tau}$ is given to list the maximum 
extension of the boundary layer section for $r_{\rm i}=0.3$. Note that the Reynolds number $Re$ for CF is given by 1145 and that the friction
Reynolds number is based on the channel half width, $Re_{\tau}=u_{\tau} L_z/(2\nu)$.}
\label{Tabpran}
\end{center}
\end{table}
\begin{figure}
\begin{center}
\includegraphics[scale=0.22]{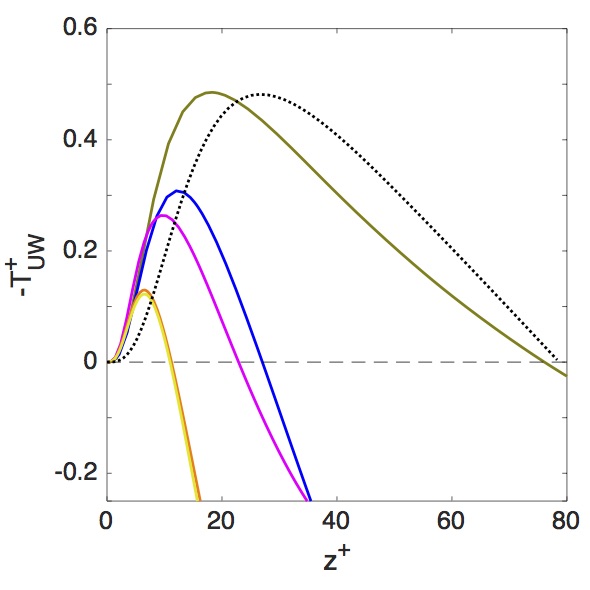}
\includegraphics[scale=0.22]{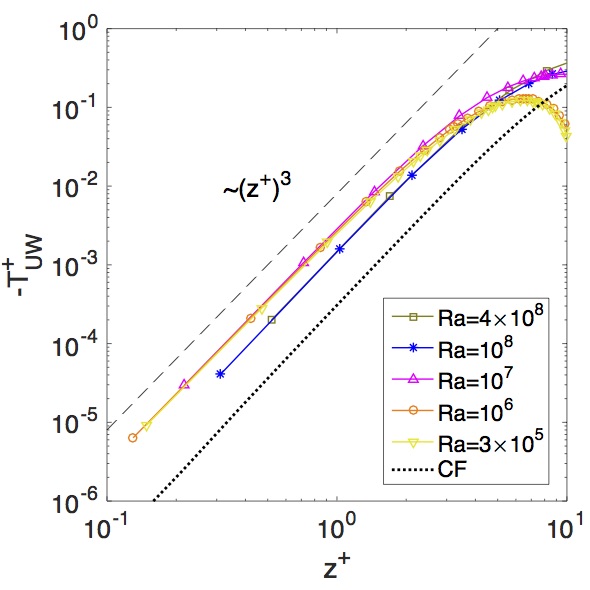}
\includegraphics[scale=0.22]{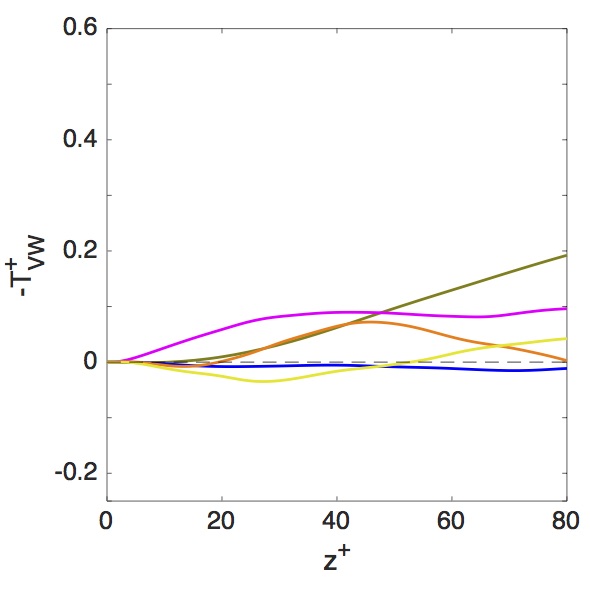}
\caption{Reynolds shear stresses versus distance from the wall. Left: stress component $T_{UW}^{+}$.  Mid: replot of $T_{UW}^{+}$
in logarithmic units. Right: stress component $T_{VW}^{+}$. The profiles are a combination of statistics from the bottom and top plates. All 
quantities are again given in inner units. The legend holds for all panels.}
\label{reynstress1}
\end{center}
\end{figure}

\subsection{Reynolds stresses}
The Reynolds shear stress, which couples the fluctuations of the streamwise velocity fluctuation to the wall-normal ones and is 
responsible for the momentum transfer from the wall into the bulk of the wall bounded flow, plays a central role for the production 
of turbulent kinetic energy. Figure \ref{reynstress1} displays the Reynolds shear stresses in the present system in inner units. 
The components are given by
\begin{align}
\label{Reyn1}
T_{UW}^{+}(z)= - \langle U^{\prime}_{\parallel}W^{\prime}_{\parallel}\rangle_{b,t}/u_{\tau}^2\,,\\
T_{VW}^{+}(z)= - \langle V^{\prime}_{\parallel}W^{\prime}_{\parallel}\rangle_{b,t}/u_{\tau}^2\,,
\end{align}
with the Reynolds decomposition
\begin{align}
\label{Reyn2}
U_{\parallel}^{\prime}(x_{\parallel},y_{\parallel},z,t)&=U_{\parallel}(x_{\parallel},y_{\parallel},z,t) - \langle U_{\parallel}(z)\rangle_{b,t}\,,\\
V_{\parallel}^{\prime}(x_{\parallel},y_{\parallel},z,t)&=V_{\parallel}(x_{\parallel},y_{\parallel},z,t) \,,\\
W_{\parallel}^{\prime}(x_{\parallel},y_{\parallel},z,t)&=u_z(x_{\parallel},y_{\parallel},z,t)-\langle u_z(z)\rangle_{b,t}\,.
\end{align}
Again, the rotation has been 
applied and the fluctuations of all three velocity components in the rotated frame have been determined subsequently. The magnitude and 
the extension from the wall into the bulk of the positive amplitudes of $T_{UW}^{+}$ are comparable  with the data of Elsnab et al. \cite{Elsnab2011}. 
In addition, the $T_{UW}^{+}$ profile for RBC5 is comparable with the channel flow run at the same $Re_{\tau}$. The maximum of the CF stress 
profile is shifted by $\Delta z^+=10$ away from the wall, the zero is almost identical. The mid panel confirms that all profiles start with a cubic $z$-dependence
from the wall. This is a consequence of the Taylor expansion in combination with the incompressibility. For example, at $z=0$ follows
\begin{align}
\label{Reyn2a}
U_{\parallel}^{\prime}(x_{\parallel},y_{\parallel},z,t)&\simeq s_x(x_{\parallel},y_{\parallel},t) z + \dots\,,\\
V_{\parallel}^{\prime}(x_{\parallel},y_{\parallel},z,t)&\simeq s_y(x_{\parallel},y_{\parallel},t) z + \dots \,,\\
W_{\parallel}^{\prime}(x_{\parallel},y_{\parallel},z,t)&\simeq -\frac{1}{2}\left(\frac{\partial s_x}{\partial x}+\frac{\partial s_y}{\partial y}\right)z^2 +\dots \,.
\end{align}
The leading order expansion coefficients are the components of the skin friction field as well as its divergence.
The vertical shift is determined by the magnitude of the shear at the plate, which is larger for all convection runs in comparison to CF (see also Section 
\ref{derivatives}).
The other Reynolds stress contribution $T_{VW}^{+}$ is indefinite for the time intervals that 
were accessible to gather statistics (see right panel of figure \ref{reynstress1}). It can be expected that this stress becomes exactly zero in the very 
long time limit.

\begin{figure}
\begin{center}
\includegraphics[scale=0.4]{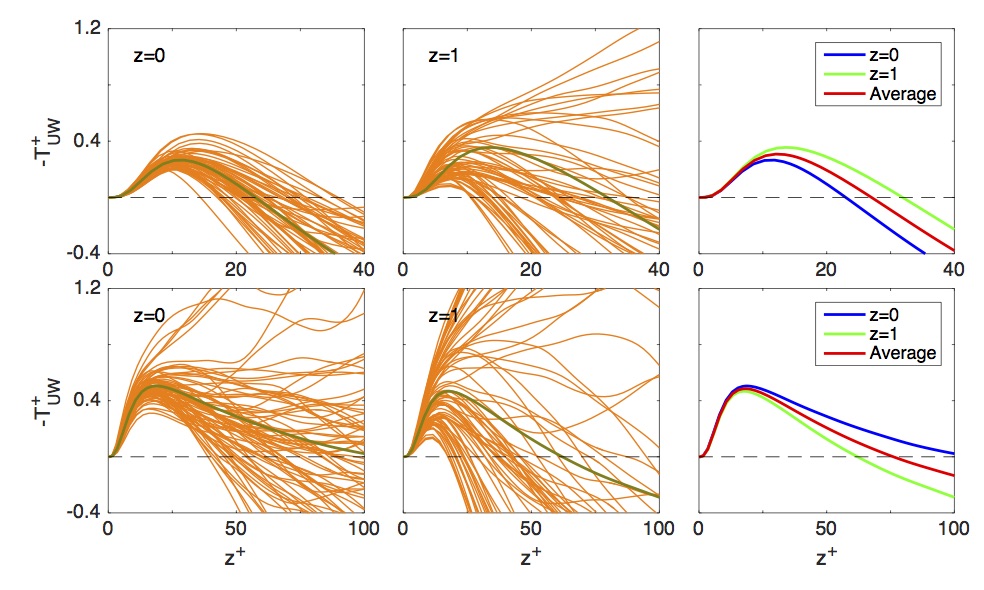}
\caption{Variability of the Reynolds shear stresses $-T^+_{UW}(z^+)$ at the bottom and top plates with respect to 
time. The data in the upper row display instantaneous profiles from 57 snapshots in orange obtained from RBC4, 
the ones in the lower row from 75 snapshots from RBC5. The corresponding average is the gray line. The two panels to
the right compare the averages the top and bottom plates as well as the total average profiles.}
\label{reynstress}
\end{center}
\end{figure}

In Figure \ref{reynstress} we show the variability in the instantaneous  Reynolds stress profiles $T_{UW}^{+}$ for RBC4 and RBC5 and at the top and 
bottom plate. The variability, particularly for RBC5 is more extreme. Note that these profiles were plotted after the system reached a statistically 
steady state, and some of the largest deviations from the average occur late in the simulation time.

\subsection{Mean temperature}
\begin{figure*}
\begin{center}
  \includegraphics[scale=0.3]{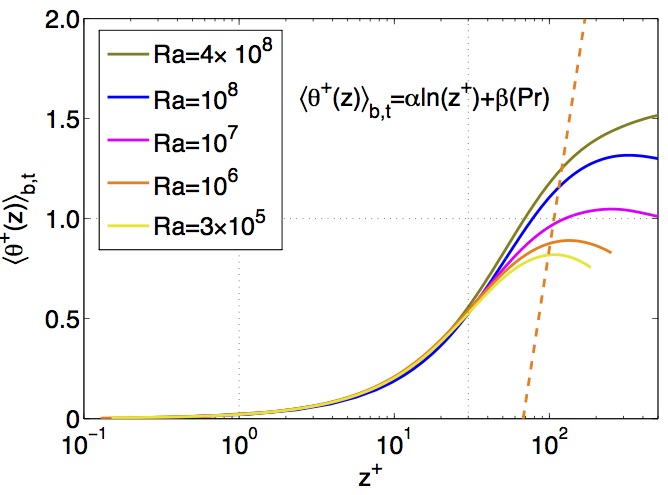}
  \includegraphics[scale=0.3]{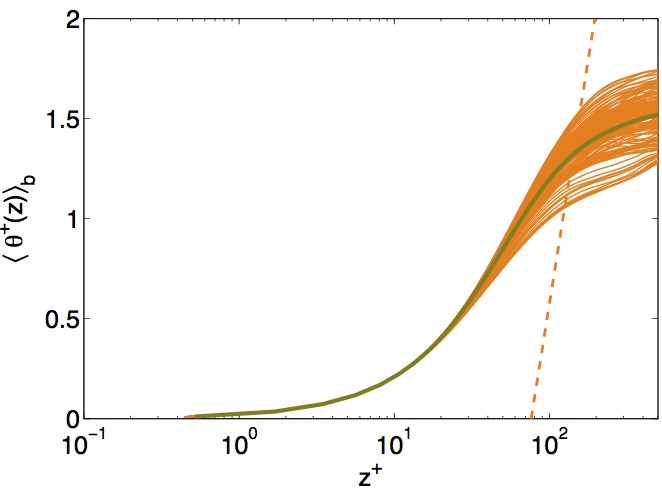}
\caption{Left: Semi-logarithmic plot of the mean temperature profiles. We display 
$\langle \theta^+(z)\rangle=\langle\langle\tilde{\theta}(z,t)\rangle_{b}/\tilde{T}_{\tau}(t)\rangle_t$ versus 
  $z^+=\langle z/z_{\tau}(t)\rangle_t$  for $Pr=0.021$. In all cases the profiles taken from the top and bottom 
plate are included in the time average. Rayleigh numbers are indicated in the figure for 
each data set. Also indicated is logarithmic law (\ref{temp3}) as a dashed line. Right: 
Temperature profiles for RBC5. The gray curve is the same as in the left. The instantanenous 
profiles are also plotted here in orange, to give a sense of the range of variation of such 
profiles with time.}
\label{temp1}
\end{center}
\end{figure*}

While the turbulent momentum transfer is significantly enhanced at low Prandtl numbers which becomes visible by the 
large Reynolds numbers, the turbulent heat transfer is strongly reduced due to the large thermal diffusivity. In this subsection, 
we will plot the mean temperature profiles in inner wall units which requires an additional quantity besides the friction velocity. 
The friction temperature, following \cite{Yaglom1979,Kader1981}, is defined by $\tilde{T}_{\tau}=-\tilde{\kappa} \tilde{u}^{-1}_{\tau}\,
\langle\partial \tilde{T}/\partial \tilde{z}|_{\tilde{z}=0}\rangle_{b,t}$. In dimensionless notation this results in 
\begin{equation}
\label{temp2}
T_{\tau}=-\Bigg\langle\frac{u_{\tau}^{-1}(t)}{\sqrt{Ra Pr}} \Bigg\langle\frac{\partial T}{\partial z}\Bigg\rangle_{b}\Bigg|_{z=0}\Bigg\rangle_t\,,
\end{equation}
where again, this quantity is evaluated at the upper plate ($z=1$) when the boundary layer at the upper plate is analyzed. 
The dimensionless temperature profile 
\begin{equation}
\label{temp3a}
\langle\theta(z)\rangle_{b,t} = \left \{ \begin{array}{ccl}
                                \dfrac{1-\langle \tilde{T}(z)\rangle_{b,t}}{\Delta \tilde{T}} & : & z\in [0,1/2]]\\
                                & & \\
                                \dfrac{\langle \tilde{T}(z)\rangle_{b,t}}{\Delta \tilde{T}}& : & z\in [1/2,1]                                                           
                                \end{array} \right.
\end{equation}
is rescaled with the dimensionless friction temperature $T_{\tau}$. Figure \ref{temp1} (left) shows 
$\langle\theta^+(z)\rangle_{b,t}=\langle\langle\theta(z,t)\rangle_{b}/T_{\tau}(t)\rangle_t$ versus 
$z^+=\langle z/z_{\tau}(t)\rangle_t$. The right panel of figure \ref{temp1} shows the range of variation in the instantaneous profiles for RBC4.
Although there is a linear range in the profiles it is not related to the temperature profiles in a turbulent boundary layer. Following Kader and 
Yaglom \cite{Yaglom1979,Kader1981} the logarithmic temperature profiles should follow
\begin{equation}
\label{temp3}
\langle\theta^+(z)\rangle= \alpha \ln z^+ + \beta(Pr)\quad\text{with}\quad \alpha\approx 2.12\,,\;\;\beta(Pr)=(3.8 Pr^{1/3}-1)^2-1+2.12 \ln Pr\,.
\end{equation}
Equation (\ref{temp3}) has been obtained by an interpolation of a comprehensive data record of turbulence experiments in pipes, channels and 
boundary layers which span a range of Prandtl numbers from 100 to 0.022.  It can be seen that the present data do not match with the Yaglom-Kader
parametrization. This could be again related to the fact that the boundary layer is not yet fully turbulent.

There is a region which is logarithmic for each temperature profile and we can fit a line to those data and find a slope. In all cases we obtain a value less than the logarithmic profile value of $\alpha \approx 2.12$ of the Yaglom-Kader parametrization. But, the slope is increasing as $Ra$ increases. This is true both instantaneously (finding the slopes of the orange curves in figure \ref{temp1} (right)) and also on average. We  provide a table which compares our instantaneous and average slopes to those of Kadar and Yaglom as well as the work done by Ahlers et. al. \cite{Ahlers2014, Wei2014}. Although our slopes are far from the Yaglom-Kader results, they are close to the results by Ahlers and co-workers, when scaled to be consistent with their units. This is noteworthy since their Rayleigh numbers were much higher ($10^{11}-10^{12}$) and this was for $Pr=0.8$ and aspect ratio $\Gamma = 1$. However, as noted in Wei and Ahlers \cite{Wei2014}, who performed experiments for $Pr=12$,  $\alpha$ decreased as  Prandtl number increased, so the trend here for our $\langle \alpha_f \rangle T_{\tau}$ values to be larger than those of Ahlers for our smaller Prandtl number is  consistent.

\begin{figure}
  \begin{center}
    \includegraphics[scale=0.3]{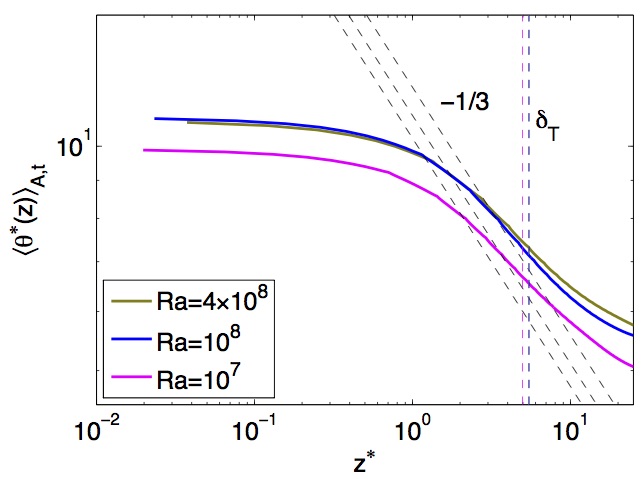}
    \includegraphics[scale=0.3]{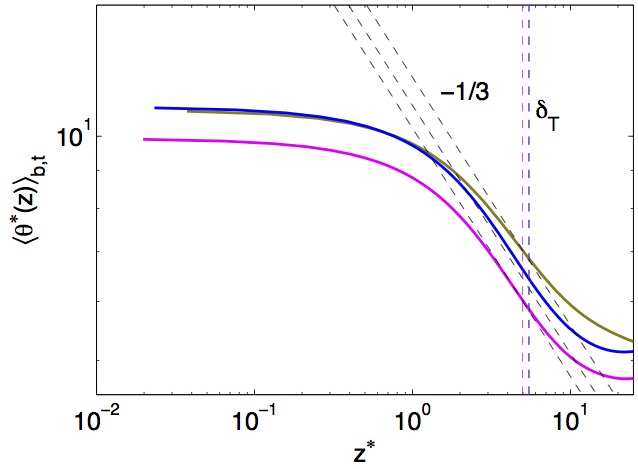}
    \caption{Test of power law behavior (\ref{tc1}) of temperature profiles.}
\label{tc1plt}
\end{center}
\end{figure}

A second rescaling was suggested by Chung et al. \cite{Chung1992} for convection at low Prandtl numbers. Following the original idea by Kraichnan \cite{Kraichnan1962} the authors developed a three-layer model consisting of a conduction layer, transition layer and a convection layer with 
corresponding characterisitic scales of length, velocity and temperature, respectively. Similar to the inner viscous units, we can take inner conductive 
units as follows. One defines a characteristic velocity scale $\tilde{u}_c=(\tilde{\kappa}^2\tilde{g}\tilde{\alpha}\tilde{Q}/\tilde{\nu})^{1/4}$, a characteristic length 
scale $\tilde{z_c}=\tilde{\kappa}/\tilde{u}_c$, and a characteristic temperature scale $\tilde{T}_c=\tilde{Q}/\tilde{u}_c$. In 
dimensionless notation this results in
\begin{equation}
\label{tc1}
z_c=\frac{1}{(Nu Ra)^{1/4}}\;\;\;\text{and}\;\;\;T_c=\left(\frac{Nu^3}{Ra}\right)^{1/4}\,.
\end{equation}
In figure \ref{tc1plt}  we replot  $\langle\theta^{\ast}(z)\rangle=\langle T(z)\rangle/T_{c}$ versus $z^{\ast}=z/z_c$. We compare the 
mean temperature profiles obtained for the whole plate and the interior section. For all three Rayleigh numbers, it is observed that the 
temperature profiles follows a -1/3 scaling with respect to $z^{\ast}=z/z_c$ in a short range for $z^{\ast}>1$, but only for the 
interior averaged profiles. Such a scaling has been predicted 
by Priestley \cite{Priestley1955} on the basis of dimensional analysis and has been detected for $Pr\gtrsim 1$ \cite{Chung1992,Mellado2012}. 
Here, we confirm the scaling for low-Prandtl-number convection.    

\section{Velocity and temperature derivatives at the plates}
\label{derivatives}
\begin{figure*}
\begin{center}
\includegraphics[scale=0.3]{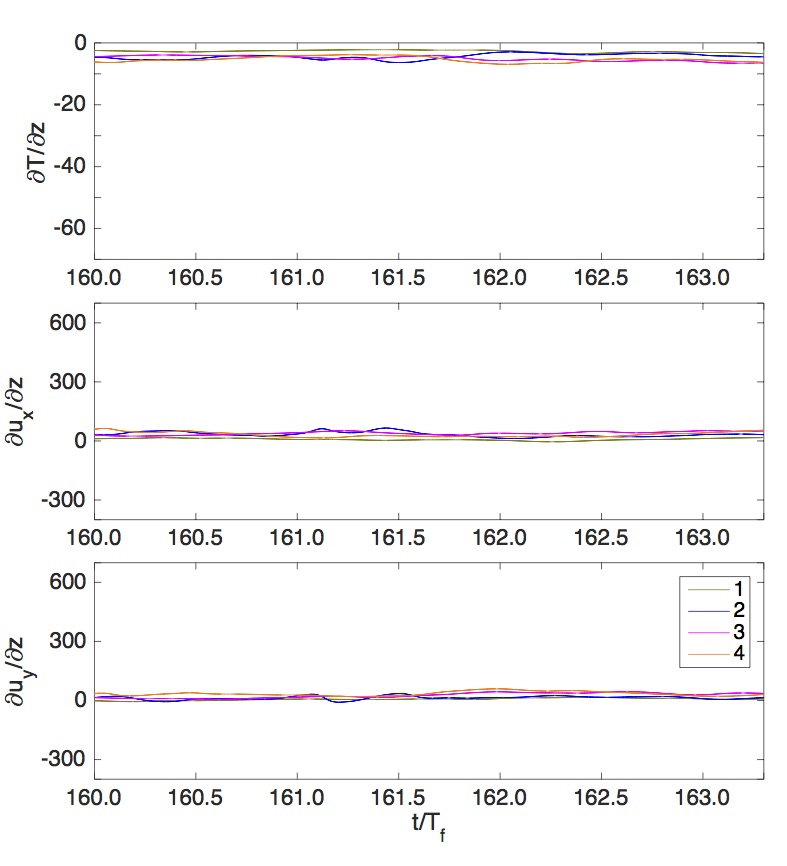}
\includegraphics[scale=0.3]{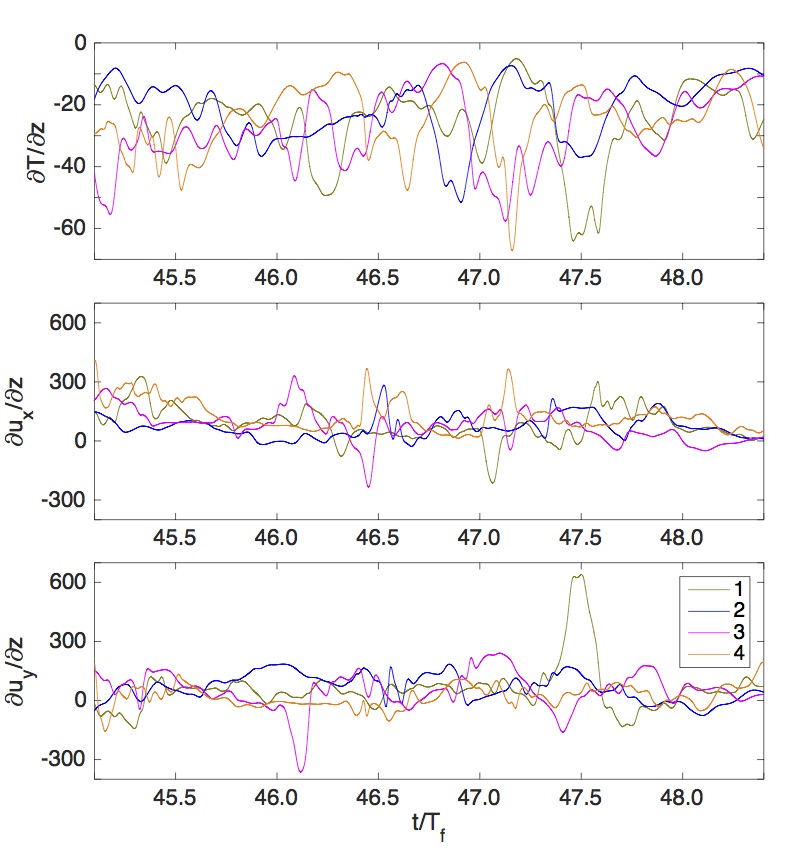}
\caption{Time series of derivatives at the plate $z=0$ which are taken for a time interval of 3.3 $T_f$ in both runs. 
The vertical temperature derivative $\partial T/\partial z$ and the two components of the skin friction field are shown. Left: 
RBC1. The time series are taken at four history points in the interior section $b$ which are approximately situated at 
$(x_1,y_1)=(0.12,-0.04)$,  $(x_2,y_2)=(-0.04,0.12)$,  $(x_3,y_3)=(-0.12,-0.04)$, and $(x_4,y_4)=(0.04,-0.12)$.  Right: 
RBC4. The four history points are $(x_1,y_1)=(0.06,0.03)$,  $(x_2,y_2)=(-0.03,0.06)$, $(x_3,y_3)=(-0.06,-0.03)$, and 
$(x_4,y_4)=(0.03,-0.06)$. The amplitudes of both data records are directly comparable in each panel.}
\label{walld}
\end{center}
\end{figure*}

It has been noted in the last section that the boundary layer structure is in some respects similar to the near-wall dynamics in planar 
wall-bounded shear flows, in particular for the simulations at the largest Rayleigh numbers. Therefore, the statistics at the wall will be studied in this last
section, and in particular the  velocity derivatives. In figure \ref{walld} the derivatives at the bottom plate, $\partial T/\partial z$, $\partial u_x/\partial z$, as well as
$\partial u_y/\partial z$ are found at each time step at four different locations (see caption of figure \ref{walld}) all in the interior section $b$ for runs RBC1 
and RBC4. The data are displayed in the same ranges over a time segment of a few free fall time units. One can clearly see that all derivatives are much 
larger and show more significant fluctuations for RBC4 than for RBC1. The time series indicate that the character of both boundary layers is already strongly 
transitional for the largest accessible Rayleigh numbers.

\begin{table}[t]
\begin{center}
\begin{tabular}{c c c c c c c}
\hline\hline
$Ra$ & $\langle \alpha_f \rangle$ &  max $\alpha_f$  & min $\alpha_f$ & $\;\;\;$Y/K$\;\;\;$ &  $\langle \alpha_f \rangle T_{\tau}$ & ABH \\ 
\hline
$3\times 10^5$ & 0.33 $\pm$ 0.01 & 0.36 & 0.28 & 2.12 & 0.22  $\pm$ 0.01  \qquad & 0.14 $\pm$ 0.02  \\
$1\times 10^6$ & 0.35 $\pm$ 0.03 & 0.41 & 0.25 & 2.12 & 0.21 $\pm$ 0.02  \qquad & 0.12 $\pm$ 0.02 \\
$1\times 10^7$ & 0.40 $\pm$ 0.05 & 0.52 & 0.28 & 2.12 & 0.20 $\pm$ 0.03 \qquad & 0.09 $\pm$ 0.01 \\
$1\times 10^8$ & 0.48 $\pm$ 0.05 & 0.59 & 0.38 & 2.12 & 0.20 $\pm$ 0.03 \qquad & 0.07 $\pm$ 0.01  \\
$4\times 10^8$ & 0.50 $\pm$ 0.04 & 0.58 & 0.38 & 2.12 & 0.18 $\pm$ 0.01 \qquad & 0.06 $\pm$ 0.01 \\
\hline
\hline
\end{tabular}  
\caption{Slopes found by fitting a log law to the region of each instantaneous  profile which follows a log law. This was done for each profile for our data set, and for the top and bottom plate. The value $\langle \alpha_f \rangle$ is the  average for each data set and the error (found by standard deviation) and the max and min values are also listed. For comparison, $\alpha$ = 2.12 by Yaglom-Kader (Y/K) is given. Finally the data from Ahlers, Bodenschatz and He \cite{Ahlers2014} (ABH) for $Pr=0.8$ is also listed, where we used their fits (either equation 4.6 or 4.7 in \cite{Ahlers2014})  then multiplied our $\langle \alpha_f \rangle$ by  $T_{\tau}$ to convert to their units.}
\label{compslope}
\end{center}
\end{table}

\begin{figure*}
\begin{center}
\includegraphics[scale=0.3]{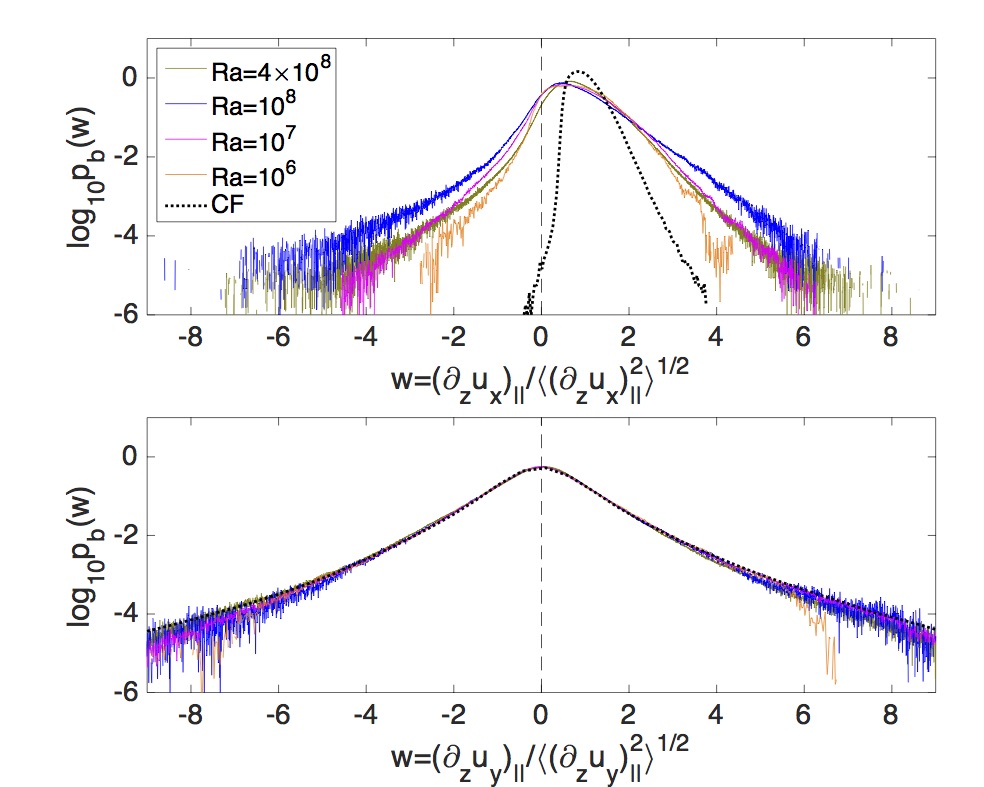}
\caption{Probability density functions of the vertical derivatives of the horizontal velocity components taken at $z=0$ and 1 for RBC2, 
RBC3, RBC4, and RBC5 in the interior plate sections. For comparison, we add the vertical derivative of the streamwise velocity component 
of the channel flow (CF) to both panels of the figure.}
\label{wall1}
\end{center}
\end{figure*}

In order to compare the velocity derivative statistics with the one in the turbulent channel with its unidirectional mean flow
we proceed as follows. Similar to the rotation (\ref{rot}) we can apply a transformation at the plates when treating the two non-vanishing velocity 
derivatives of the velocity gradient tensor as a two-dimensional vector field. Therefore the definition (\ref{angle}) is adapted to
\begin{equation}
\label{angle1}
\langle {\gamma(t)}\rangle_{b} = \arctan \left [\frac{\langle \partial u_y/\partial z(z=0,t)\rangle_{b}}{\langle \partial u_x/\partial z(z=0,t)\rangle_{b}} \right]\,,
\end{equation}
and the original plane-by-plane transformation $\hat{R}_3(z,t)$ is changed to the rotation $\hat{R}_2(z=0)$  
\begin{equation}
  \label{rot1}
  \left(\begin{array}{c}
                                \partial_z u_x(t)|_{z=0} \\
                                \partial_z u_y(t)|_{z=0}
                                \end{array}\right)_{\parallel} = 
   \left(\begin{array}{cc}
                                \cos \langle \gamma(t)\rangle_{b}  & \sin \langle \gamma(t)\rangle_{b}  \\
                                -\sin \langle \gamma(t)\rangle_{b}  & \cos \langle \gamma(t)\rangle_{b}
                                \end{array}\right) 
  \left(\begin{array}{c}
                                \partial_z u_x(t)|_{z=0} \\
                                \partial_z u_y(t)|_{z=0}                              
                                \end{array}\right) \;.
\end{equation}
The same transformation $\hat{R}_2(z=1)$ follows for the top plate at $z=1$ with a corresponding angle. 
In figure \ref{wall1}, PDFs of $(\partial u_x/\partial z)_{\parallel}$ and $(\partial u_y/\partial z)_{\parallel}$ are shown for RBC2, RBC3, RBC4, and RBC5
along with CF, the channel flow run at a comparable Rayleigh number, each of which is scaled by its respective root mean square (rms) value. 
Note the symmetry for $(\partial u_{y}/\partial z)_{\parallel}$, for all runs,  further supporting that our transformation to a streamwise and 
spanwise direction makes the present data better comparable to a channel setup. Conversely the 
pdfs for  $(\partial u_x/\partial z)_{\parallel}$ are asymmetric, indicating a net shear flow for $U_{\parallel}$. The pdfs become wider as the Rayleigh 
number increases, indicating an increase of the intermittent fluctuations of the derivatives for increasing $Ra$. Their shape agrees remarkably 
well with the findings of Lenaers et al. \cite{Lenaers2014} (see e.g. their figure 2). The increasingly wider tails for the present data underlines 
an increasingly transitional character of the viscous boundary layer.

While the PDF of the streamwise velocity derivative of CF contains a small negative tail only, the distributions of both
components of the skin friction field for RBC have large tails for both negative and positive values. Thus it is expected that a significant number of critical 
points exists, i.e., points at $(x,y,z=0)$ and $(x,y,z=1)$ at which ${\bm s}=0$. The following pairs of complex eigenvalues $\lambda_k=a_k+i b_k$ are possible:  
saddle points with $\lambda_1=a_1<0$ and $\lambda_2=a_2>0$; unstable nodes with $\lambda_1=a_1>0$ and $\lambda_2=a_2>0$ as well as
stable nodes with $\lambda_1=a_1<0$ and $\lambda_2=a_2<0$. Also possible are unstable foci with $\lambda_{1,2}=a\pm ib$ 
or stable foci with $\lambda_{1,2}=-a\pm ib$ both of which with $a>0$ \cite{Chong2012,Bandaru2015}.

\begin{figure*}
\begin{center}
\includegraphics[scale=0.15]{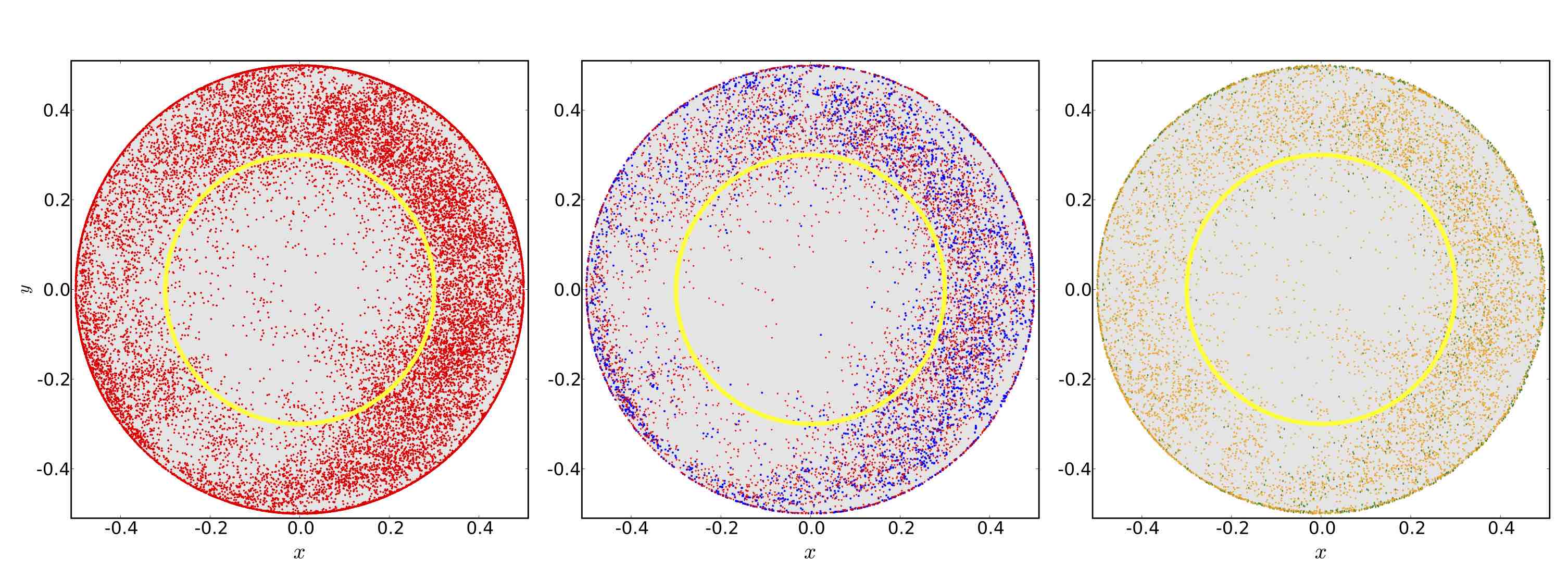}
\caption{Location of critical points for the skin friction field at the bottom plate $z=0$ for run RBC4  for 57 snapshots spanning 6.7 free-fall times. 
The left panel shows the location of the saddle points, the middle panel shows  the location of the stable nodes (red) and foci (blue) and 
the right panel shows the unstable nodes (orange) and foci (green). The yellow circle is the inner radius  $r=r_{\text i}$. The view onto the 
plate is from above}.
\label{critical}
\end{center}
\end{figure*}

The dynamics in the boundary layer of a turbulent convection flow can be quantified by computing these critical points of the skin 
friction field at the bottom or top plate as in \cite{Bandaru2015}.  For example, saddle points or stable foci can be associated with plume emission 
and unstable nodes or foci can be associated with plumes hitting the plate. Figure \ref{critical} shows that this mainly occurs near the outer region  
of the plate. If we confine ourselves to be inside the region defined by the yellow circle in figure \ref{critical}, there are fewer critical points and the 
region is thus more similar to the near-wall region in a wall--bounded shear flow \cite{Chong2012,Lenaers2014}. This holds in particular if one combines 
all three panels of figure 
\ref{critical}.

\section{Summary and conclusions}
The structure of the boundary layers in a high-Reynolds number turbulent Rayleigh-B\'{e}nard flow has been studied 
from the perspective of transitional wall-bounded flows, such as a channel flow. Since the momentum transfer response (and thus
the large-scale Reynolds number) is very large in liquid metal convection flows at very low Prandtl numbers compared to air or water,
the viscous boundary layer fluctuates particularly strongly which is quantified by Reynolds shear stress profiles and the statistics of 
derivatives at the walls. Our analysis is based on a series of three-dimensional direct numerical simulation runs for $Pr=0.021$. 
The high spectral resolution allowed us to study the derivative statistics and to determine friction velocities and temperatures.

The torsion and the varying orientation of the large-scale circulation in the closed cylindrical cell is (partly) removed by a symmetry transformation
that is applied for each grid plane between bottom and top separately. It is then shown that the mean streamwise velocity approaches the 
standard logarithmic law of the wall from below. This is in contrast to a transient low-Reynolds number channel which would approach a 
logarithmic scaling from above caused by the parabolic laminar flow profile at very small Reynolds numbers. 

When the sidewall effects are excluded, the temperature profiles come close to a power law scaling similar to Chung et al. \cite{Chung1992}. 
Although we could also fit a logarithmic law to the profiles, the slope differs significantly from what is expected for a turbulent boundary layer \cite{Yaglom1979,Kader1981}. It remains to be seen if this scaling 
changes when the Prandtl number is even further decreased and/or the Rayleigh number is further increased. 

The Reynolds shear stress component $T_{UW}^+$, which is obtained at the same $Re_{\tau}$ for runs RBC5 and CF, obeys qualitatively 
the same shape although the maxima are shifted by 10 wall units with respect to each other.  Together with the profiles which have been
obtained for $T_{VW}^+$, this demonstrates that the transformation (\ref{rot}) can identify a streamwise direction and thus effectively remove a significant part
of the complex three-dimensional mean flow structure. 

\begin{figure*}
\begin{center}
\includegraphics[scale=0.12]{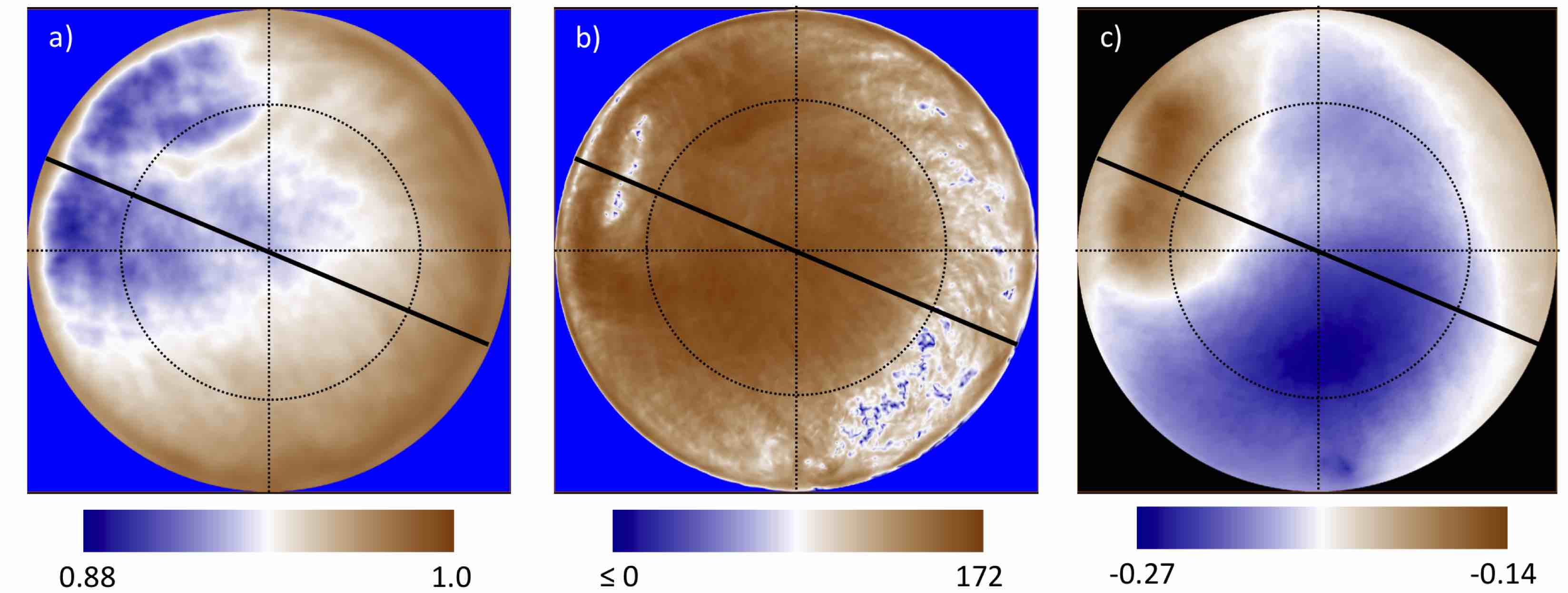}
\caption{Time-averaged boundary layer structure at the bottom plate of the turbulent convection flow RBC4 
at $Pr=0.021$ and $Ra=10^8$. (a) Temperature $T$ at $z=0.0024$ which corresponds with $0.09 \delta_T$. (b) 
Magnitude of skin friction $|{\bm s}|$ in logarithmic units at $z=0$. (c) Pressure $p$ at  $z=0.0024$. The view on the 
bottom plate is from below. All time averages are taken over 6.72 $T_f$. The dotted horizontal and vertical lines are a guide 
to the eye. The dotted circle indicates the interior plate section with $r\le r_i$. The solid thick line indicates the time-averaged 
mean flow orientation (upper left to lower right) which is taken over the interior section.}
\label{Fig14}
\end{center}
\end{figure*}

How can the structure of the boundary layer be described on average? We go back to figure \ref{Fig1} at the beginning and replot in figure
\ref{Fig14} the time-averaged slice cuts of temperature, skin friction magnitude and kinematic pressure taken at the same heights as in figure \ref{Fig1}. 
The solid line in all three panels indicates the mean orientation of the flow in the vicinity of the plate. It is obtained again by averaging over the interior 
plate section. The following picture arises:
\begin{itemize}
\item The plume impact region at the bottom plate is on average colder than the rest of the plate region. Temperature increases along the mean 
streamwise direction. The hotter plate region is where the LSC rises up towards the top plate -- the plume ejection region.
\item The skin friction field magnitude shows the biggest spatial variability in the impact and ejection regions. 
This is also where most of the critical points, ${\bm s}=0$, are observed.
As shown in figure \ref{critical}, the majority of these points are found outside the interior plate section. The skin friction magnitude is largest
in the interior section where the LSC sweeps across the plates and generates strong shear.
\item The interior plate section is well approximated by a favorable pressure gradient boundary layer. A local pressure maximum is clearly associated 
with the plume impact. Pressure increases again slightly further downstream at the opposite edge of the interior region. This might be connected with 
the increase of temperature in the vicinity of the side wall. 
\item With increasing Rayleigh number, it is found that the wall stress (or skin friction) field components fluctuate increasingly stronger
which also underlines the increasingly transitional character of the viscous boundary layer.  
\end{itemize}
This is a general and coarse-grained picture which is mostly related to the viscous boundary layer dynamics. Our DNS record allows to extrapolate
the existing data in order to predict when a friction Reynolds number $Re_{\tau}\sim 200$ is obtained that results in a turbulent channel flow as discussed
in the landmark paper by Kim, Moin and Moser \cite{Kim1987}. Our present low-$Pr$ data suggest turbulence inside the viscous boundary layer for $Ra\gtrsim 10^{11}$. This value would be consistent with the experiments by Glazier et al. \cite{Glazier1999} that went up to Rayleigh numbers 
of $Ra\sim 10^{11}$. However, their cell for the highest $Ra$ had an aspect ratio $\Gamma=1/2$ which reduces the downstream evolution length 
$2r_{\rm i}/z_{\tau}$ at a given Rayleigh number (see table \ref{Tabpran}) and thus the scale over which the boundary layer can become turbulent.

As a next step, it would be interesting to study  the near-wall structure formation to more detail. These investigations are already in 
progress and will be reported elsewhere. Another interesting direction is to lower the Prandtl number even further, e.g. to values 
$Pr < 10^{-2}$ which are typical for liquid sodium \cite{Scheel2016}. Numerical simulations at larger Rayleigh numbers in sodium 
at $Pr=0.005$ are also currently underway and will be discussed in the near future.

\acknowledgements
AP and JDS were partly supported by the Research Training Group GRK 1567 on Lorentz Force Velocimetry which is funded 
by the Deutsche Forschungsgemeinschaft. We acknowledge supercomputing time at the J\"ulich Supercomputing Centre 
which was provided by the Large Scale Project HIL09 and the Scientific Big Data Analytics Project SBDA003 of the John von 
Neumann Institute for Computing. Furthermore, we 
acknowledge an award of computer time provided by the INCITE program. This research used resources of the Argonne Leadership 
Computing Facility at  Argonne National Laboratory, which is supported by the DOE under contract DE-AC02-06CH11357. We are 
indebted to Sergei I. Chernyshenko, Joseph Klewicki and Katepalli R. Sreenivasan for their helpful comments.

\end{document}